\documentclass[12pt,preprint]{aastex}
\usepackage{graphicx}

\def\ec{$\eta$~Car}

\def\hst{{\it HST}}

\def\vlt{{\it VLT}}

\def\kms{km~s$^{-1}$}
\def\den{cm$^{-3}$}

\def\lm{$\lambda$}

\newcommand{\Msun}{\hbox{$M_\odot$}}

\newcommand{\altion}[2]{\textup{#1}\,\textsc{#2}}

\begin{document}

\title{The variable 6307\AA\ emission line in the spectrum of Eta Carinae: blueshifted [S III] $\lambda$6313 from the interacting winds}

\shorttitle{$\lambda$6307 Identified}
\shortauthors{Gull et al.}

\author{ T.~R. Gull\altaffilmark{1}}

\altaffiltext{1}{Laboratory for Extrasolar Planets and Stellar Astrophysics, Exploration of the Universe Division, Code 667, Goddard Space Flight Center, Greenbelt, MD 20771}

\begin{abstract}The 6307\AA\ emission line in the spectrum of \ec\ \citep{Martin06a} is blue-shifted [\altion{S}{iii}] \lm6313 emission originating from  the outer wind structures of the massive binary system. We realized the identification while analyzing multiple forbidden emission lines not normally seen in the spectra of massive stars. The high spatial and moderate spectral resolutions of \hst/STIS resolve forbidden lines of Fe$^+$, N$^+$, Fe$^{+2}$, S$^{+2}$, Ne$^{+2}$ and Ar$^{+2}$ into spatially and velocity-resolved  rope-like features originating from collisionally-excited ions photo-ionized by UV  photons or collisions.
While the [\altion{Fe}{ii}] emission extends across  a velocity range of $\pm$500~\kms\ out to 0\farcs7,
 more highly ionized forbidden emissions ( [\altion{N}{ii}], [\altion{Fe}{iii}], [\altion{S}{iii}], [\altion{Ar}{iii}], and [\altion{Ne}{iii}]) range in velocity from $-$500 to $+$200 \kms, but spatially extend outward to only  0\farcs4. The [\altion{Fe}{ii}] defines the outer regions of the massive primary wind. The [\altion{N}{ii}], [\altion{Fe}{iii}] emission define the  the outer wind interaction regions directly photo-ionized by far-UV radiation. Variations in emission of [\altion{S}{iii}] \lm\lm 9533, 9071 and 6313 suggest density ranges of 10$^6$ - 10$^{10}$~\den\ for electron temperatures ranging from 8,000 to 13,000\degr\ K. Mapping the temporal changes of the emission structure at critical phases of the 5.54-year period  will provide important diagnostics of the interacting winds.\end{abstract}


\section{Introduction} \citet{Martin06a} found  a variable emission line  centered at 6307\AA\ in multiple spectra of \ec, recorded by \hst/STIS and by \vlt/UVES. 
 They were unable to identify the origin of the emission line, but demonstrated that the line was present across the high state (defined by presence of forbidden lines of doubly-ionized elements, see \cite{Damineli08a} and references therein) and disappeared during the low state. \\
 \indent Recently \citet{Gull09a}, using the same spectra, focused on the spatially-extended forbidden line emission both from high-ionization (herein defined as $>$14 eV) and low-ionization (8-13 eV). We found that the forbidden emission originated from 1)  the Weigelt condensations \citep{Weigelt86},  as narrow lines  centered on $-$43 \kms, 2)  the boundaries of the primary wind (\ec~A) as rope-like [\altion{Fe}{ii}], photo-ionized by mid-UV and collisionally excited at densities around N$_c$=10$^7$\den, and 3) the wind interaction region, as high-ionization emission  from [\altion{N}{ii}], [\altion{Fe}{iii}], [\altion{Ar}{iii}], [\altion{Ne}{iii}] and [\altion{S}{iii}], photo-ionized by far-UV and collisionally excited for densities, n$_c$ ranging from 10$^5$ to 10$^8$\den.  The high-ionization emission lines are present both for the Weigelt condensations and the wind interaction region during the 5 year high state, but every 5.5 years, disappear during the low state. X-ray models \citep{Pittard02, Parkin09a} place the binary periastron event near the onset of the low state with a three to six month recovery.\\
\indent We applied a 3D SPH (smoothed particle hydrodynamics) model \citep{Okazaki08a} extended out to 1700 AU (0\farcs67) to match the spatial structure seen in these forbidden lines and realized that the bulk of the high-ionization emission structure originated from the wind interaction region in the outer portion of the massive wind structure. We  found that portions of the wind interaction structure,  moving ballistically outward, are directly illuminated by the far-UV radiation of the hot secondary, \ec~B, leading to  highly-ionized, collisionally excited gas and hence the high-ionization extended emission.\\
\indent Further examination of the \hst/STIS longslit spectra showed that the previously unidentified emission at 6307\AA\ is  blue-shifted [\altion{S}{iii}] $\lambda$6313 emission from the interacting wind structure. The evidence is subtle, but convincing. We summarize the observations in Section 2. A description of how the spectro-images were produced is  in Section 3. The forbidden emission structures are described in Section 4. Discussion in Section 5 provides insight on the ionization and excitation leading to [\altion{S}{iii}] emission and  the potential for monitoring changes with orbital phase, including mapping temperature with density dependence. We conclude with a summary in Section 6. Throughout this paper, all wavelengths are  in vacuum, the velocities are heliocentric, directions are compass points (N=north, NNW= north by northwest and the phase of the binary orbit is referenced to the X-ray minimum beginning at 1997.9604 \citep{Corcoran05a}. 
\vfill\eject
\section{The \hst/STIS Observations}
\begin{figure}
   \includegraphics[angle=0,width=3.4in]{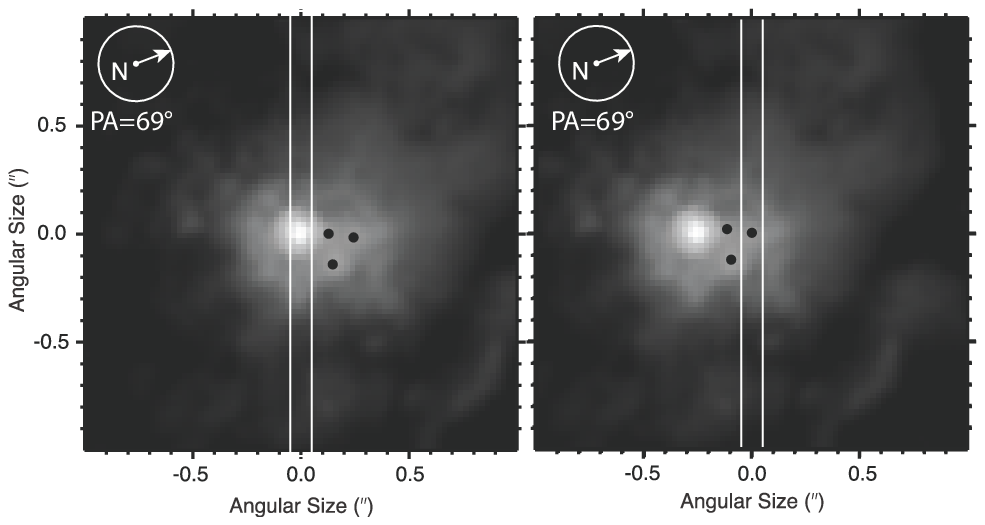}
 \caption{{\bf \hst/STIS aperture positions: Left:} Centered on \ec.  {\bf Right:} Centered on Weigelt D. The 2\arcsec$\times$2\arcsec\ Field of View is extracted from an \hst/ACS image recorded in February 2003 through the 550M filter. Weigelt condensations B, C and D are indicated by the black dots. The projection of the 52\arcsec$\times$0\farcs1 aperture is indicated. Note the field is rotated by 69\degr\ placing the aperture vertical. North is indicated by the compass. \label{Aperture} }
 \end{figure}
The spectra discussed here are a portion of the Eta Carinae Treasury observations accessible through the STScI archives ( http://archive.stsci.edu/prepds/etacar) as reduced by a special reduction tool developed by K. Ishibashi and K. Davidson. For brevity we focus on two sets of observations, recorded in July 2002 ($\phi$=0.820) and July 2003 ($\phi$=1.001).  \\
\indent Observations were recorded with the \hst/STIS moderate dispersion gratings and CCD detector through the 52\arcsec$\times$0\farcs1 aperture. We wanted to monitor the change in both \ec\ and the Weigelt condensations \citep{Weigelt86}, which drop in excitation during the low state. However the range in aperture position angle (PA) is limited by the required orientation of  \hst\ solar panels and changes throughout the year. 
can prevent inclusion of any one of the three Weigelt condensations within the aperture when centered on \ec\ (Figure \ref{Aperture}). 
For observations centered around the X-ray minimum, predicted to be around 1 July, 2003, we scheduled a visit one year earlier, July 2, 2002 (orbital phase, $\phi$=0.820), at a pre-selected PA=69\degr, that would be accessible just before($\phi$=0.995) and after the X-ray minimum ($\phi$=1.001). During all three visits, separate observations centered on \ec\ and Weigelt D were obtained with the aperture placed as shown in Figure \ref{Aperture}.   Additional information on the observations are presented in \citet{Martin06a} and \citet{ Gull09a}.

\section{Spatially-resolved Emission}
\begin{figure}
\epsscale{.8}
\plotone{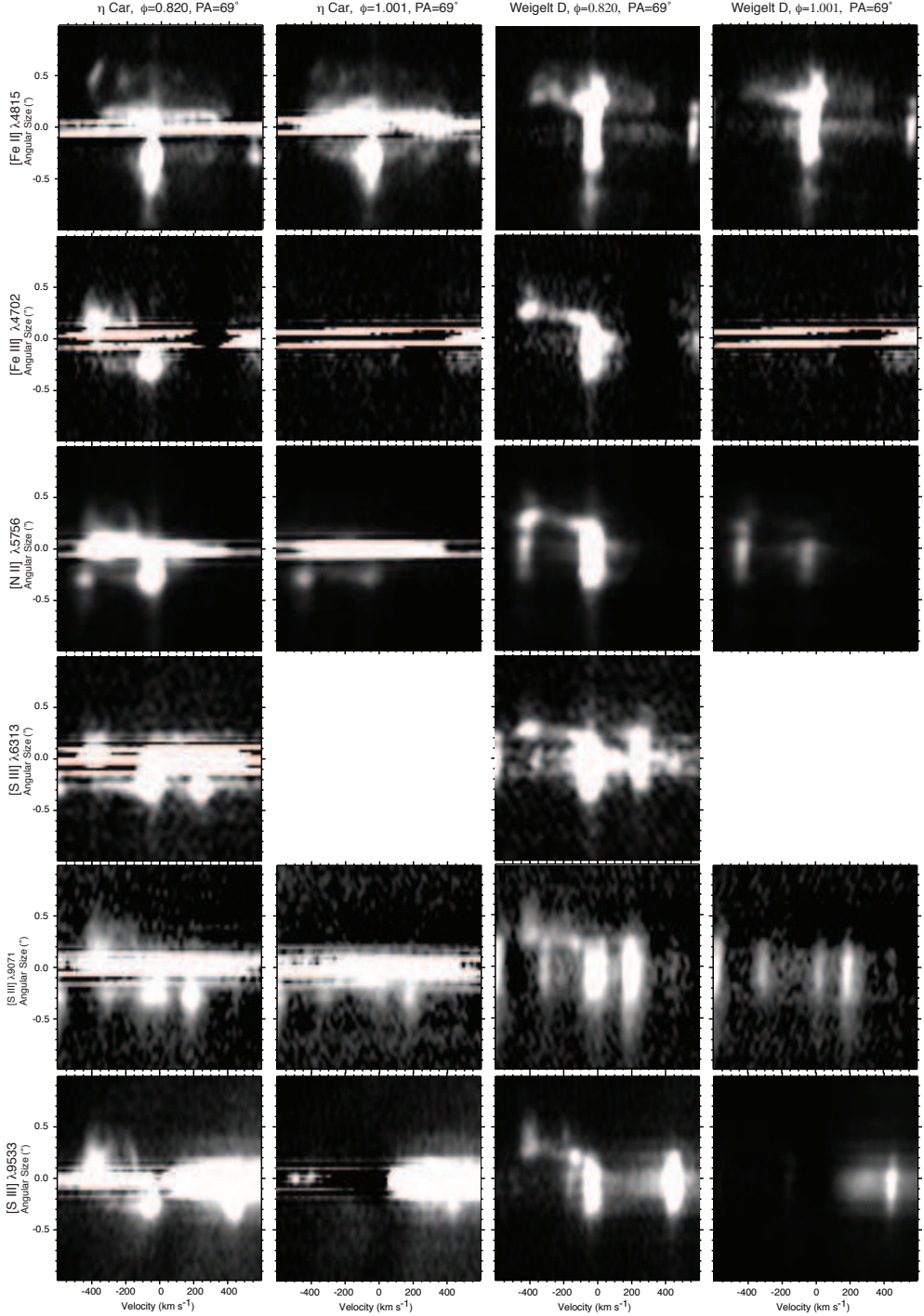} 
\caption{{\bf Spectro-images}  centered on \ec\ (Columns 1 and 2) and on Weigelt D (Columns 3 and 4).  {\bf[\altion{Fe}{II}]~$\lambda$4815} (Row 1) {\bf [\altion{Fe}{III}]~$\lambda$4703} (Row 2) 
  {\bf[\altion{N}{II}]~$\lambda$5756} (Row 3) and
 {\bf [\altion{S}{III}]~$\lambda\lambda$6313, 9071, 9533} (Rows 4$-$6) 
 \label{Spectro-images}
 {\bf Note:} Continuum in regions free of narrow or broad-line emission has been subtracted on a spatial line-by-line basis to display the extended emission structure. All plots are with a grey level proportional to $\sqrt{Intensity}$. All observations were recorded with PA=69\degr.}
   \end{figure}

The \hst/STIS spatial-resolution (0\farcs1 at H$\alpha$) separates the spectrum of \ec's core from  extended structures, especially the narrow line emission that originates from the Weigelt condensations \citep{Davidson95},
 located between 0\farcs1 to 0\farcs3 in the NW quadrant relative to  \ec\ (see Figure \ref{Aperture}).  As described by \citet{Gull09a}, we  found considerable differences between many broad forbidden emission line profiles of \ec\ as recorded by the \vlt/UVES and the \hst/STIS. Extractions with a 0\farcs127-high slice of the STIS spectra (five half rows in the reduced spectro-images) yielded broad wind line profiles for \altion{H}{i}, \altion{He}{I} and \altion{Fe}{ii} lines that compared favorably with those recorded by \vlt/UVES. In contrast,  broad profiles of forbidden lines recorded by \vlt/UVES were nearly absent in the \hst/STIS extractions centered on \ec. Examination of the \hst/STIS long aperture spectra revealed faint structure in these emission lines  extending out to 0\farcs7. However, the observed  emission was highly variable both with aperture PA  and   orbital phase, $\phi$. Clearly the forbidden line emission is spatially extended on scales resolved by \hst/STIS but not by \vlt/UVES.\\
\indent We examined individual lines in more detail and attempted to enhance visibility of the extended line emission by several reduction procedures. While various spatial filters were tried, the best results were obtained by subtraction of measured continuum on a spatial row-by-row basis. We used  spectral plots 
of the Weigelt condensations  \citep{Zethson01b} to identify 10 to 20\AA\ intervals  with no obvious presence of narrow  or  broad line emission. At each position along the aperture, we measured and subtracted the average continuum in that spectral interval. Examples of the resulting spectro-images (intensity images with x$=$velocity and y$=$angular size)  are presented in Figure \ref{Spectro-images}.\\
\clearpage
\clearpage
 \indent These spectro-images provide only a qualitative view of the line profile. We caution the reader that quantitative measures require much more precision for the following reasons: \begin{enumerate}\item The STIS utilized a three-axis mechanism to select a grating and  to set the correct tilt angle for the spectral interval of choice. While return to that grating position is within a few CCD pixels, variations in the tilt are not fully reproducible. A tilt of 1/20 pixel along the 1024 element row led to significant photometric variation when attempting to extract spectra at the 0\farcs1 spatial resolution. \item  The standard calibration for the STIS photometry is properly referenced to extractions of a stellar spectrum with a 2\arcsec-wide aperture, allowing for full capture of flux from a point source. Apertures with widths comparable to the diffraction limit of \hst\ sample the point spread function of the telescope, which changes dynamically even within an orbit. \item Charge transfer inefficiency (CTI) leads to a trail in the direction of columns and, with on-orbit time, increases. For complex sources like extended structures, a proper extraction is not available. \end{enumerate} These problems complicate attempts to show extended structure in the vicinity of a bright star, which is exactly the situation with \ec. This leads to the obvious  linear striations at the star position in the spectrally-dispersed (velocity) coordinate. These variations  do affect  measures of the extended emission closest to the stellar position. However, for offsets to Weigelt D,  the stellar flux is blocked by the aperture, and  quantitative measures are then possible. We  note  that the  stellar spectrum scattered from the direction of Weigelt D is very different from the direct spectrum of \ec. The lack of P Cygni absorption  in H$\alpha$\ across the high state indicates fully-ionized hydrogen in the region spatially located between \ec\ and Weigelt D \citep{Gull09a}. On the side of caution, we  limit this discussion to a description of the spatial and velocity structure of the lines. Even with qualitative descriptions, we gain much insight  on the spatially resolved wind interactions and the source of the 6307\AA\ emission.\\
 \section{Description of the emission structures}
\indent We refer the reader to Figure \ref{Spectro-images} for the following descriptions of the forbidden line emission.  The first two columns of spectro-images are extracted from spectra centered on \ec\  in the high state ($\phi$=0.820) and early in the low state ($\phi$=1.001). Likewise, columns 3 and 4 are centered on Weigelt D (A more complete summary on variation of the emission structures with ionization potential and orbital phase is presented by \citet{Gull09a}).\\
\indent Four basic structures contribute to these spectro-images: 
\begin{enumerate} \item The central core of \ec, not resolved by \hst\ at 0\farcs1, which contributes the bulk of the continuum and P Cygni wind lines, notably of \altion{H}{i} and \altion{Fe}{ii}. \item Weigelt D and other, lesser condensations that contribute many narrow  emission lines centered at $-$40 \kms. \item Rope-like structures of high-ionization forbidden emission lines with velocity components extending from $+$200 to $-$500 \kms. \item Noticeably more-diffuse, rope-like structures of low-ionization forbidden emission lines, specifically [\altion{Fe}{ii}].\end{enumerate}
Spectro-images of [\altion{Fe}{ii}] $\lambda$4815 (Row 1) show narrow rope-like features extending to 0\farcs7 at $-$500 \kms and other, more diffuse structures closer to the star extending $\pm$500 \kms. The narrow emission at $-$40 \kms\ originates from extended structure WSW of \ec, not noted by \citet{Weigelt86}, but present throughout the observational period from 1999 to 2004 whenever the STIS aperture sampled this position. The narrow [\altion{Fe}{ii}] emission centered on Weigelt D extends 0\farcs5 E and W of Weigelt D, but at about 0\farcs25 E of D, a diffuse emission extends to $-$400 \kms and away from \ec. During the low state, the outer [\altion{Fe}{ii}] emission drops, becomes more diffuse and is located closer to \ec.\\
\indent The  [\altion{Fe}{iii}]~$\lambda$4703 (Row 2)  is interior to the rope-like [\altion{Fe}{ii}] $\lambda$4815. A series of highly filamentary  loops extend from  \ec\ to the east at  velocities from $-$40 to $-$500 \kms. No red-shifted velocity components are seen at this PA. However, \citet{Gull09a} find that at PA=$-$28\degr, observed six times from 1998.0 to 2004.3 ($\phi$=0.000 to 1.122),  red-shifted components extend to $+$200 \kms\  at early phases, but fade late in the high state. All [\altion{Fe}{iii}]~$\lambda$4703 disappears during the low state.\\
\indent The [\altion{N}{ii}]~$\lambda$5756 (Row 3) has very similar structure to that of [\altion{Fe}{iii}] $\lambda$4703 with higher S/N. A  narrow emission line, [\altion{Fe}{ii}] $\lambda$5748, appears at the $-$550 \kms\ position and persists in the low state, along with weak [\altion{N}{ii}] $\lambda$5756. The structure of [\altion{N}{ii}] $\lambda$5756 extends from $-$40 to $-500$ \kms\ in the spectro-image centered on Weigelt D during the high state, but also disappears in the low state.\\
\indent Three [\altion{S}{iii}] lines are shown in Rows 4$-$6 as each is important in accounting for the 6307\AA\ emission.  Unfortunately, the  [\altion{S}{iii}] $\lambda$6313 (Row 4) was recorded only at $\phi$=0.820, but the other two lines were observed at both phases.  Most noticeable in the spectro-image of \ec\ at $\phi$=0.820 is a knot of emission, centered on the stellar position at $-$400 \kms. The effective wavelength is  6307\AA.  That spectral interval was recorded at other phases, but at other position angles, during the 2003.5 minimum with no  [\altion{S}{iii}] $\lambda$6313 present either at the positions of \ec\ or Weigelt D. The extended structure is less apparent in the spectro-image of \ec\ at $\phi$=0.820, but is well-defined in the spectro-image centered on Weigelt D. Narrow lines of [\altion{O}{i}] $\lambda$6302, \altion{Fe}{ii} $\lambda$$\lambda$6307, 6309 and 6319 contaminate the spectro-image and persist during the minimum.\\
\indent The   [\altion{S}{iii}] $\lambda$ 9071 (Row 5) weak emission extends off of \ec, but several narrow  lines (\altion{N}{i} $\lambda$9063,  \altion{Fe}{ii} $\lambda$9073) also contribute to the spectro-image. The high-velocity arc of [\altion{S}{iii}] $\lambda$ 9071 extends blueward from Weigelt D. \\
\indent The  [\altion{S}{iii}] $\lambda$9533 emission (Row 6) is quite similar to that of [\altion{Fe}{ii}] $\lambda$4703 (Row 2) and confirms that   [\altion{S}{iii}] emission extends to $-$500 \kms. The bright emission to the red of [\altion{S}{iii}] $\lambda$9533 is \altion{H}{i}~Pa~8 $\lambda$9548, which originates primarily from the central core. \\
\section{Discussion}
\indent We associate the low-ionization structure with the massive, slow-moving wind of \ec~A. The high-ionization emission is from the interacting wind region piled up by the fast-moving, less-massive wind of \ec~B \citep{Pittard02}. The bulk of the interacting wind, by its velocity, appears to be mostly ionized wind of \ec~A. The higher velocity side of the shock is likely less dense and more highly ionized by \ec~B.\\ 
\indent Using the 3D SPH models of \citet{Okazaki08a} and simple geometric models, we determined that the high-ionization emission originates from a distorted paraboloidal structure lying in the skirt of the Homunculus. Based upon the blue-shifted velocities and near symmetry for PAs ranging from $+$22 to $+$38\degr, the paraboloid points in our general direction with axis of rotation projecting onto the sky at PA$\approx-$25\degr.  \\
\indent \citet{Martin06a} performed a very complete analysis on the unidentified 6307\AA\ line  in both the \hst/STIS and \vlt/UVES spectra, finding very similar behavior with orbital phase. Their search of possible line identifications focused primarily on singly-ionized species such as Fe$^+$, V$^+$ and S$^+$, although they do list [\altion{S}{iii}] as a narrow nebular line identified by \citet{Zethson01b} in the spectrum of the Weigelt condensations. Their candidate of greatest interest appeared to be \altion{Fe}{iii}] $\lambda$6306.43 with unknown atomic data for the transition.\\
\indent Most important in their analysis was the tracking of the strength of the emission throughout the 5.5-year orbit. They found that the line disappeared during the low state, but might be  anti-correlated with \altion{Fe}{ii} $\lambda$5529. Both suggest a high-ionization source. \citet{Nielsen07b} analyzed the behavior of the \altion{He}{i} absorption, finding an anti-correlation with \altion{Fe}{ii} absorption.\\
\indent Salient are three facts: \begin{enumerate} \item On the star, both \hst/STIS and \vlt/UVES see the same emission bump with similar strengths. \item The emission  correlates with high-ionization variations, not the behavior of the low-ionization emission of \altion{Fe}{ii}. \item The extended emission of the extended emission correlated with [\altion{S}{iii}] correlates very well with the extended emission identified with [\altion{S}{iii}] $\lambda\lambda$9071 and 9533.\\\end{enumerate}
\section{Conclusions}
We have presented conclusive evidence that the emission line at 6307\AA, noted in the spectra of \ec\ by \citet{Martin06a} is blue-shifted emission of [\altion{S}{iii}] \lm6313 originating from the distorted paraboloidal, interaction region located between the massive binary members. While the massive primary, \ec~A, provides the dominant wind ejecta, 10$^{-3}$ \Msun/y at 500 \kms, the hotter secondary provides a less massive, faster wind, 10$^{-5}$ \Msun/y at 3000 \kms, and far-UV photons that ionize iron, neon, argon and sulfur to doubly ionized states. Thermal collisions, mid-UV photons and possible charge exchange excite the doubly-ionized species to upper states with forbidden transitions leading to forbidden line emission in regions with densities close to n$_c$. Specifically, [\altion{S}{iii}] \lm\lm9533, 9071 and 6313 have extended spatial structures. The intensity ratio (Flux (\lm9533) + Flux (\lm9071)/Flux(\lm6313) leads to density estimates ranging from 10$^7$ - 10$^8$ \den on the scale of 0\farcs1, the limit of \hst/STIS spatial capabilities. Mapping in these and other doubly-ionized lines will provide powerful measures for models of wind interactions using various 3-D hydrodynamical codes.
 \acknowledgements
The observations were accomplished with the NASA/ESA Hubble Space Telescope. Support for Program numbers  7302, 8036, 8483, 8619, 9083, 9337, 9420, 9973, 10957 and 11273 was provided by NASA directly to the Space Telescope Imaging Spectrograph Science Team and through  grants from the Space Telescope Science Institute, which is operated by the Association of Universities for Research in Astronomy, Incorporated, under NASA contract NAS5-26555.  All analysis was done using STIS IDT software tools on data available through the \hst\ \ec\ Treasury public archive.
\bibliographystyle{apj}
\bibliography{ms} 

\end{document}